\documentclass[12pt]{aastex7}
\hypersetup{linkcolor=red,citecolor=blue,filecolor=cyan,urlcolor=magenta}
\usepackage{verbatim}  

\newcommand{\Lya}     {Ly$\alpha$}    
\newcommand{\Lyb}     {Ly$\beta$}     

\newcommand {\HI}      {\ion{H}{1}}    
\newcommand {\HII}     {\ion{H}{2}}    

\newcommand {\HeI}     {\ion{He}{1}}   
\newcommand {\HeII}    {\ion{He}{2}}   
\newcommand {\HeIII}   {\ion{He}{3}}   

\newcommand {\OI}      {\ion{O}{1}}   
\newcommand {\OII}     {\ion{O}{2}}   
\newcommand {\OIII}    {\ion{O}{3}}   
\newcommand {\OIV}    {\ion{O}{4}}    
\newcommand {\OV}     {\ion{O}{5}}    
\newcommand {\OVI}    {\ion{O}{6}}    

\newcommand {\CII}     {\ion{C}{2}}    
\newcommand {\CIII}    {\ion{C}{3}}    
\newcommand {\CIV}    {\ion{C}{4}}

\newcommand {\NIII}     {\ion{N}{3}}
\newcommand {\NIV}     {\ion{N}{4}}

\newcommand {\MgII}     {\ion{Mg}{2}}  
 
\newcommand {\MgX}     {\ion{Mg}{10}}  

\newcommand {\AlIII}  {\ion{Al}{3}}    

\newcommand {\SiIV}   {\ion{Si}{4}}
\newcommand {\SiIII}  {\ion{Si}{3}}
\newcommand {\SiII}   {\ion{Si}{2}}

\newcommand {\NeIV}    {\ion{Ne}{4}}    
\newcommand {\NeV}     {\ion{Ne}{5}}    
\newcommand {\NeVI}    {\ion{Ne}{6}}    
\newcommand {\NeVIII}  {\ion{Ne}{8}}    

\newcommand {\FeII}   {\ion{Fe}{2}}

\newcommand {\kms}    {km~s$^{-1}$}

\newcommand {\etal}   {et~al.}

\shorttitle{QSO EUV continuum}
\shortauthors{Shull et al.}

\begin{document}

\title{Quasar Spectral Energy Distributions in the Rest-Frame EUV: \\
Hubble-COS Spectra of Two Ultra-luminous Quasars}

\email{michael.shull@colorado.edu}\tabletypesize{\footnotesize}

\author[0000-0002-4594-9936]{J. Michael Shull}\email{michael.shull@colorado.edu}
\affiliation{CASA, Department of Astrophysical \& Planetary Sciences, University of Colorado, Boulder, CO 80309, USA; }
\affiliation{Department of Physics and Astronomy, University of North Carolina, Chapel Hill, NC 27599, USA; }

\author[0000-0002-3120-7173]{Rongmon Bordoloi}\email{rbordol@ncsu.edu}
\affiliation{Department of Physics, North Carolina State University, Raleigh, NC 27695 USA;}

\author[0000-0003-4738-6601]{Charles Danforth}\email{charles.danforth@colorado.edu}
\affiliation{CASA, Department of Astrophysical \& Planetary Sciences, University of Colorado, Boulder CO 80309, USA;}

\begin{abstract}
Using the Cosmic Origins Spectrograph (COS) aboard the {\it Hubble Space Telescope} with
both far-UV (FUV) and near-UV (NUV) gratings, we measure the ionizing spectra of two 
bright, intermediate-redshift quasars in their rest-frame extreme ultraviolet (EUV).
The availability of both NUV and FUV spectra allows us to define the quasar continuum
and correct for strong Lyman-limit systems (LLS) that fall in the gap between the
FUV and optical.  Each AGN has a prominent LLS, but the flux recovery shortward of their 
Lyman edges allows us to fit and restore the true AGN continuum. In the EUV (450--912~\AA) 
these AGN have flux distributions, $F_{\nu} \propto \nu^{-\alpha_{\nu}}$, with spectral 
indices $\alpha_{\nu} = 1.11\pm0.22$ (SBS~1010+535, $z_{\rm AGN} = 1.5086$) and 
$\alpha_{\nu} = 0.98\pm0.22$ (HS~0747+4259, $z_{\rm AGN} = 1.9006$), both considerably 
harder than the mean index, $\alpha_{\nu} = 1.41\pm0.15$, in a COS composite spectrum 
of 159 UV-bright AGN.  These two AGN are outliers in the index distribution, perhaps 
resulting from their extremely high UV luminosity ($10^{48}~{\rm erg~s}^{-1}$), 
estimated black-hole masses (0.5--1)$\times10^{10}~M_{\odot}$, and
effects on the inner accretion disk and Comptonized winds.  
 
\end{abstract}

\keywords{}

\section{Introduction} \label{sec:intro}
Despite several decades of ultraviolet (UV) observations, the ionizing spectrum of active galactic 
nuclei (AGN) remains uncertain, but not for the lack of trying \citep{Zheng1997,Telfer2002,Scott2004,Shull2012,Stevans2014,Lusso2015,Tilton2016}.   
Photons in the AGN rest-frame extreme ultraviolet (EUV, 100--912~\AA) 
photoionize gas in galactic halos, circumgalactic medium (CGM), and intergalactic medium (IGM) and 
provide diagnostics of the AGN accretion disk and broad emission-line region.  Unfortunately, direct
observations of the rest-frame FUV/EUV shortward of 1216~\AA\ are difficult at $z > 2$, owing to 
strong \Lya-forest absorption in the IGM.  However, the \Lya\ forest at $z \approx$ 0.5--2 is thinner, 
enabling observations of the redshifted EUV toward intermediate-redshift quasars with UV
spectrographs aboard the {\it Far Ultraviolet Spectroscopic Explorer} (FUSE) and the 
{\it Hubble Space Telescope} (HST).   The EUV also contains resonance lines of \HeI, the 
Li-like doublets of \NeVIII\  and \MgX, and a range of ionization stages of oxygen 
(\OII, \OIII, \OIV, \OV, \OVI), nitrogen (\NIII, \NIV), and neon (\NeIV, \NeV, \NeVI, \NeVIII).  

\medskip

 A significant number of bright AGN at $z > 1$ have been studied with the Cosmic Origins Spectrograph 
 (COS) on HST, using the moderate resolution (20 \kms) gratings (G130M and G160M).
However, the number of observed targets at $z > 1.5$ remains small, with a limited sample of AGN 
that probe rest-frame wavelengths below 600~\AA.  This paper continues our long-term program of
measuring the ionizing spectra of quasars.  \cite{Shull2012} presented a 
composite UV spectrum using 22 AGN ($0.026 \leq z_{\rm AGN} \leq 1.44$) observed with the COS 
moderate resolution gratings (G130M and G160M). \cite{Stevans2014} extended this survey to 
159 AGN at $0.001 < z_{\rm AGN}\leq 1.476$.  For somewhat fainter quasars, \cite{Tilton2016} 
obtained spectra for 11 AGN at $1.45 \leq z_{\rm AGN} \leq 2.14$, using the COS/140L 
grating at resolving power $R \approx$ 1500--4000 from 1130--2150~\AA.  

\medskip

However, with only far-ultraviolet (FUV) spectra (1130--2000~\AA), the COS gratings (G140L, G130M, 
G160M) sometimes fail to measure the curvature in AGN continuum slope, and they occasionally 
miss strong \HI\ absorption systems just longward of 2000~\AA, in the near-ultraviolet (NUV).  
The flux recovery of these Lyman limit systems (LLS) shortward of their 912~\AA\ edges can 
masquerade as a steep spectral slope in the FUV, as was the case with the quasar  
SBS~1010+535 \citep{Tilton2016}. Thus, the AGN continuum slope can be uncertain 
when extrapolating from the FUV into the EUV.  These uncertainties and other selection factors may
contribute to the range of spectral indices found in previous AGN composite spectra ({\bf Table 1}).  

\medskip

Our previous AGN composite projects fitted the underlying continuum flux distributions with 
power-law shapes, $F_{\nu} \propto \nu^{-\alpha_{\nu}}$.  After de-reddening the observed AGN 
spectra, we placed the continuum below broad emission lines in the rest-frame FUV and EUV and 
above narrow interstellar and intergalactic absorption lines.   We also corrected for strong 
absorbers:  LLS, partial LLS, and occasional damped \Lya\ absorbers.  \cite{Stevans2014} presented 
a composite UV spectrum of 159 AGN, finding a gradual break at AGN rest-frame wavelength 
$\sim$1000~\AA\ and a steepening in spectral index from $\alpha_{\nu} = 0.83\pm0.09$ 
(1200--2000~\AA) to $\alpha_{\nu} = 1.41\pm0.15$ (500--1000~\AA).   That composite and a 
lower-resolution study of 11 quasars with COS G140L data \citep{Tilton2016} highlighted the 
benefits of obtaining NUV spectra to fill the gap between FUV and optical wavelengths and
provide a broader wavelength interval, anchored to sections of continuum largely free of broad 
emission lines.  At intermediate AGN redshifts, this ``NUV gap"  contains useful absorption-line 
diagnostics, including \HI\ and \HeI\ absorption lines from the LLS that can alter the continuum.  

\medskip

This paper investigates the ionizing continua of the two FUV-brightest intermediate-redshift 
quasars from our previous HST programs.   Our latest Cycle-30 HST program (PID-17071) was 
designed to bridge the NUV gap with COS/G230L spectra at observed wavelengths 1710--3550~\AA.  
The rest-frame EUV continuum was previously observed with COS/G140L in HST Cycle 21 (PID-13302) 
and Cycle 25 (PID-15084).  Our previous HST/COS composite survey \citep{Tilton2016} used 
UV-bright AGN at intermediate redshifts ($1.5 \leq z_{\rm AGN} \leq 2.1$) selected on the 
basis of their high GALEX fluxes \citep{Bianchi2014}.  Our more recent Cycle-30 sample of 
12 AGN included eight AGN at $z \geq 1.7$, with G140L spectra probing rest-frame wavelengths 
down to 360--420~\AA.  The two AGN analyzed in this paper had the greatest far-UV fluxes and 
were observed with full NUV wavelength coverage, using three settings of the COS/G230L 
grating combined with G140L and optical spectra.  
 
 \medskip
 
 In Section 2, we describe the HST/COS observations of these two intermediate AGN and derive
 properties of their LLS.  We explain our methods for fitting the underlying continuum, after
 de-reddening and correcting for \HI\ photoelectric absorption shortward of the LLS edges 
 (at rest-frame 912~\AA).  We then compare the EUV continuum slopes with the mean values 
 found in previous AGN composite spectra.  The two quasars discussed here have harder (flatter) 
 continua from $\lambda_{\rm rest} = 912$~\AA\ down to 450~\AA, compared to most earlier composite 
 spectra.  These AGN are ultra-luminous, with $\lambda L_{\lambda} \approx 10^{48}$ erg~s$^{-1}$ at 
 $\lambda \approx$ 1350~\AA\ and are outliers in the distribution of EUV spectral indices 
 \citep{Stevans2014}.  We summarize our results in Section 3 and suggest the need for additional 
 studies of AGN Lyman continua and hot gas in the galactic halos associated with these LLS.  
 
\section{Observations}

{\bf Table 2} describes our two targets with details on the COS observations. The optical spectra 
(3800--9500~\AA) were obtained from archives of the Sloan Digital Sky Survey (SDSS) for 10 of the 
12 targets in our full survey.  Two southern targets were observed with Magellan (courtesy of 
Dr.\ Fakhri Zahedy).  Here, we present observations of the two brightest AGN in our sample 
(SBS~1010+535 at $z_{\rm AGN} = 1.509$ and HS~0747+4259 at $z_{\rm AGN} = 1.901$) with full FUV 
and NUV wavelength coverage.  Both sight lines exhibit a prominent LLS in or near the NUV gap, at 
absorber redshifts $z_a = 1.269$ (observed at $\lambda_{\rm LLS} = 2069$~\AA) and $z_a = 1.0784$ 
(at $\lambda_{\rm LLS} = 1897$~\AA) respectively.  With FUV/NUV spectra coverage, we were able to 
correct for the LLS absorption and establish an accurate AGN continuum.  

\medskip

The ultraviolet spectral data presented in this article were obtained from the Mikulski Archive for 
Space Telescopes (MAST) at the Space Telescope Science Institute. The specific observations analyzed 
can be accessed via  \dataset[doi:10.17909/ppq0-be75]{https://doi.org/10.17909/ppq0-be75}.
The FUV data were taken with COS/G140L in a single central wavelength (CENWAVE) setting (1105~\AA).
The NUV data were taken with COS/G230L in three settings (2365~\AA, 3000~\AA, 3360~\AA)
as listed in Table 2. Together, these gratings provide full UV wavelength coverage with good S/N 
from 1130~\AA\ to 3100~\AA.  The SDSS spectra extended from 3800~\AA\ to 9500~\AA.  
We downloaded \texttt{x1d} files from the HST MAST archive, processed using the latest version 
of the \textsc{calcos} pipeline. The observations were conducted in single visits (at different times) 
with G140L and G230L.  SBS~1010+535 was observed with 2 orbits (G140L) and 3 orbits (G230L).  
HS~0747+4259 was observed with 1 orbit (G140L) and 3 orbits (G230L).  All observations used four 
FP-split exposures, each with durations of 550--650~s.

\medskip

For each grating, individual \texttt{x1d} files were co-added using the Hubble Advanced Spectral 
Products (HASP) pipeline \citep{HASP2024}, which performs automated, visit-level co-addition by 
aligning spectra in wavelength space, filtering based on flux consistency and data quality flags, 
and summing fluxes with inverse-variance weighting. Prior to co-addition, each \texttt{x1d} file 
was visually inspected to verify consistency in the wavelength solution across all exposures. 
For each target, data from individual gratings 
were combined separately, producing a single co-added spectrum per grating.
We then combined the final G140L (FUV) and G230L (NUV) spectra using inverse-variance weighting to 
construct the final science-ready spectra for each target. The two gratings overlap in wavelength 
between 1750~\AA\ and 1950~\AA.  In this region, the G140L data were resampled to the lower resolution 
of G230L before being combined.  The optical spectra were obtained from the SDSS--DR16 quasar 
catalog\footnote{The SDSS catalog is at 
\url{https://www.sdss4.org/dr17/algorithms/qso_catalog}.  SBS1010+535 was observed in the 
SDSS-Legacy program and listed with $z = 1.50861\pm0.00020$ and magnitude $u = 16.53\pm0.01$
(data file = spec-0304-52381-0275.fits).  HS~0747+4259 was observed in the SDSS-BOSS 
program and listed at $z = 1.89865\pm0.00015$ and $u = 16.13\pm0.01$ 
(data file = spec-366955481-0006.fits).}.  

\medskip

{\bf Figure 1} shows the final combined FUV, NUV, and optical spectra, plotted in the AGN rest 
frame.  Even in the low-resolution G140L/G230L spectra, one can detect curvature in the 
slope of the underlying AGN continuum, as well as small undulations, broad emission lines, 
and numerous absorption features from the interstellar medium and LLSs.  Details of the flux 
alignment between FUV, NUV, and optical spectra are described below, in Section 2.2. 
Redshifts for the AGN and the LLS were found by cross-correlation of emission/absorption features. 
Our values of $z_{\rm AGN}$ differ slightly from those in the SDSS catalog. 
The SDSS spectra are labeled with the prominent (rest-frame UV) emission lines, 
including \CIV\ (1548~\AA), \CIII] (1909~\AA), \MgII\ (2800~\AA), and the \SiIV--\OIV] blend 
(1397, 1404~\AA).  The COS spectra are labeled with positions of the two LLS, observed at 
2069~\AA\ (SBS~1010+535) and 1897~\AA\ (HS~0747+4259).  We also label prominent 
AGN broad emission lines  (\NeVIII, \OIV, \Lyb+\OVI), and the location of the 
Lyman continuum (LyC) at 912~\AA\ (AGN rest-frame).  
No absorption is apparent at the ionization edges of \HI\ (912~\AA) or \HeI\ (504~\AA) in
either AGN, with optical depth limits $\tau_{\rm HI} < 0.1$ and $\tau_{\rm HeI} < 0.1$.  This 
suggests low column densities of \HI\ and \HeI\ intrinsic to the AGN and a LyC escape 
fraction of nearly 100\%.  We also detected absorption lines from the Galactic interstellar 
medium and the LLS  (e.g., \HI, \HeI, \CIII, \SiIII, \OVI). However, the low S/N and low 
resolution prevented us from making accurate measurements of their strengths.

\medskip

Because they probe the high end of the AGN UV luminosity function, the 12 AGN in our 
Cycle-30 study are not a complete or unbiased sample.  They represent the best  COS 
observations of known bright AGNs at redshifts $z_{\rm AGN} \approx$ 1.5--2.1, and 
they were chosen without prior knowledge of their EUV spectral shapes.  The two AGN 
in this paper have high UV luminosities, $\lambda L_{\lambda}$, estimated from fluxes
$F_{\lambda} \approx 4 \times10^{-15}$ erg~cm$^{-2}$~s$^{-1}$ \AA$^{-1}$
at rest-frame FUV wavelengths $\lambda \approx 1100-1500$~\AA,
\begin{equation}
\lambda L_{\lambda} = (1.0\times10^{48}~{\rm erg~s}^{-1})\; h_{70}^2
 \left[ \frac{d_L}{40~{\rm Gpc}} \right]^2 
\left[ \frac{F_{\lambda}}{4\times10^{-15}~{\rm erg~cm}^{-2}~{\rm s}^{-1}~{\rm \AA}^{-1}} \right]
\left[ \frac{\lambda}{1350~{\rm \AA}} \right] \; .
\end{equation}
 For this luminosity estimate, we converted the specific flux $F_{\lambda}$ at 
 rest-frame wavelength $\lambda \approx 1350$~\AA\ to monochromatic luminosity 
 $L_{\lambda} = 4 \pi d_L^2 F_{\lambda}$.  We adopted cosmological luminosity 
 distances $d_L(z)= 36.2$~Gpc ($z = 1.509$) and 48.2~Gpc ($z = 1.901$) computed for 
 a flat $\Lambda$CDM universe with $H_0 = (70~{\rm km~s}^{-1}~{\rm Mpc}^{-1})h_{70}$, 
 $\Omega_m= 0.315$, and $\Omega_{\Lambda} = 0.685$.  Evidently, both quasars are
 ultra-luminous, with $\lambda L_{\lambda} \approx 10^{48}$ ergs~s$^{-1}$, 
 suggesting that they were observed during periods of high mass-accretion rates.
 We estimated the black hole masses from calibrated relations with AGN 
 line widths (FWHM) and monochromatic luminosities ($\lambda L_{\lambda}$), 
\begin{equation}
    \log \left[ \frac {M_{\rm BH}}{M_{\odot}} \right] = \alpha 
     + \beta \log  \left[ \frac {{\rm FWHM}}{1000~{\rm km~s}^{-1}} \right]
     + \gamma \log \left[ \frac {\lambda L_{\lambda}}{10^{44}~{\rm erg~s}^{-1}} \right] 
\end{equation}
For \CIV\ $\lambda 1549$, we use parameters $\alpha = 7.48\pm0.24$, $\beta = 0.52\pm0.07$, 
and $\gamma = 0.56\pm0.48$ from \cite{Park2013} and monochromatic luminosity at 1350~\AA.  
For \MgII\ $\lambda 2800$, we use parameters $\alpha = 7.13\pm0.27$, $\beta = 0.5$, and 
$\gamma = 1.51\pm0.49$ from \cite{Wang2009} and luminosities at 3000~\AA.  
From SDSS spectra of HS~0747+4259, we obtained measurements of both \CIV\ 
(FWHM = $7140\pm100$~\kms\ and 1350~\AA\ luminosity $9.4\times10^{47}~{\rm erg~s}^{-1}$) 
and \MgII\ (FWHM = $3450\pm100$~\kms\ and 3000~\AA\ luminosity 
$7.5\times10^{47}~{\rm erg~s}^{-1}$).  In both quasars, the estimated black hole masses
are $M_{\rm BH} \approx$ (0.5--1)$\times 10^{10}~M_{\odot}$, with propagated errors of 0.55 dex.  
Their luminosities are close to the Eddington limit, placing them in a 
special class of AGN compared to those used to construct previous FUV/EUV composite spectra. 

\subsection{Lyman Limit Systems} 

Both AGN spectra exhibit a prominent LLS, defined as a strong quasar absorption system 
with \HI\ column density $N_{\rm HI} \geq 1.6\times10^{17}~{\rm cm}^{-2}$, optically thick 
at the Lyman continuum edge (911.75~\AA).  These systems are nearly always accompanied
by heavy-element absorption lines with a range of metallicities.  In a LLS sample
at $z \leq 1$, \cite{Lehner2013} found metallicity peaks at $\log (Z/Z_{\odot}) \approx -1.6$ 
(3\% solar) and  $-0.3$ (50\% solar), interpreted as low-metallicity gas (inflow or accretion) 
and metal-enriched gas (outflows from galactic winds) respectively.  The LLS spectra often include 
singly ionized species (\MgII, \SiII, \FeII, \CII) and intermediate ions (\CIII, \CIV, \SiIII, \SiIV).  
Many have associated hot halos, given their active star-formation. 
 
 \medskip
 
The two LLS studied here have moderate optical depths ($\tau_{\rm LL}$) at the Lyman
limit (LL), related to the \HI\ column density $N_{\mathrm{H\,I}}$ (in cm$^{-2}$) by 
$\tau_{\rm LL} = (6.304\times10^{-18}~{\rm cm}^2) N_{\rm HI}$.
The first quasar SBS~1010+535 ($z_{\rm AGN} = 1.5086$) has a LLS at $z_a = 1.269$ 
with $\tau_{\rm LL} = 1.70\pm0.08$.  The quasar HS~0747+4259 ($z_{\rm AGN} = 1.9006$) 
has a LLS  at $z_a = 1.0784$ with $\tau_{\rm LL} = 1.04\pm0.06$. These measurements were performed 
by modeling the continuum decrement $\tau_{\rm LL}$ at the Lyman-limit break and inferring 
$N_{\rm HI}$ from the optical depth using an MCMC fitting method implemented in the 
\texttt{rbcodes} package \citep{rbcodes}. Such LLS with significant flux recovery in the 
FUV are rare, but they can provide valuable diagnostics of halo gas, 
with \HI\ column densities easily inferred from the LL optical depths.  
For the LLS toward SBS~1010+535, we find $\log N_{\mathrm{H\,I}} = 17.431\pm0.021$.
For the LLS toward HS~0747+4259, we find $\log N_{\mathrm{H\,I}} = 17.218\pm0.024$.
These \HI\ column densities are ideal for rest-frame EUV (450--900~\AA) spectroscopy:
high enough to detect metals, but sufficiently low ($17.2 < \log N_{\rm HI}  < 17.5)$ 
for significant flux recovery shortward of their Lyman edges. 
Interestingly, \cite{Reimers2006} obtained high-resolution HST/STIS spectra of one of our
quasars (HS~0747+4259), listing 16 \OVI\ absorption systems between $1.07 \leq z_a < 1.87$.
One of them (system \#19 at $z_a = 1.0778$) is almost certainly the LLS that we measured at
$z_{\rm LLS} = 1.0784$.  However, they did not provide its column densities, 
other than noting the presence of \Lya\ and several metal lines
(\CII\ 1334, \SiII\ 1260, \SiIII\ 1206, \AlIII\ 1854, \CIV\ 1548, 1550, 
\SiIV\ 1393, 1402, and \OVI\ 1031, 1037).  

\subsection{Rest-Frame EUV Continua} 

{\bf Figure~2} shows our fits to the FUV and NUV continua of both AGN, extending 
from rest-frame wavelengths 1200~\AA\ down to 450~\AA. The absorption-corrected EUV continua 
were obtained by multiplying the observed flux by $\exp(\tau_{\lambda})$ to account for the
optical depth $\tau_{\lambda}$ of \ion{H}{1} absorption (see eqs.\ [3] and [4] below). 
The AGN continuum in the EUV was fitted using rest-frame wavelength bands generally free of 
strong broad emission lines. As in our previous composite spectra, these include windows at 
528--532~\AA, 660--670~\AA, 715--735~\AA, and 855--880~\AA.  Additional line-free windows 
contributed to the FUV fitting. To determine the slope of the quasar continua using HST/COS 
spectra, we implemented a six-step analysis pipeline: 
\begin{enumerate}

    \item Correct for Galactic extinction using the \cite{Fitzpatrick1999} reddening law with 
    $R_V = 3.1$ and $E(B-V)$ values inferred from the Planck \citep{Planck2016} optical depth 
    map at 353~GHz, converted via $E(B-V) = 1.49 \times 10^4 \, \tau_{353}$.
   
    \item Shift spectra to the AGN rest frame using measured redshifts.
    
    \item Combine the COS (NUV and FUV) spectra with SDSS optical spectra to better constrain the 
    longer-wavelength power-law slope. Because of known discrepancies in absolute flux calibration 
    between HST/COS and SDSS spectra, we renormalized the SDSS spectra by constant factors: 
    0.8 for SBS~1010+535 and 0.75 for HS~0747+4259.  We saw no evidence for flux variability
    in the different epochs of the FUV and NUV observations.  
    
    \item Correct for Lyman-limit absorption by restoring the absorbed flux with 
    $\tau_{\lambda} = \sigma_{\lambda} N_{\rm HI}$ (see Figure 2).
    
    \item Identify continuum windows free of strong AGN emission lines at specific rest-frame 
    wavelength ranges: 485-495~\AA, 528–532~\AA, 585–595~\AA, 660–670~\AA, 715–735~\AA, 
    855–880~\AA, 1090–1105~\AA, 1430–1480~\AA, 1700-1780~\AA, and 2670–2690~\AA\ and apply 
    ``sigma clipping" to the flux values within each window to remove outliers, such as residual 
    emission or absorption features.  Sigma clipping excludes any flux values that deviate 
    by more than 3$\sigma$ from the median. The median of the clipped flux was then adopted as 
    the continuum anchor for the power-law fitting. Uncertainties on the continuum flux were 
    estimated via bootstrap resampling.
    
    \item Fit a broken power law to the selected continuum regions using Markov Chain Monte 
    Carlo (MCMC) sampling with the \texttt{emcee} package \citep{Foreman_Mackey_2013}.
    
\end{enumerate}

To de-redden the quasar spectra, we adopted selective extinction values inferred from 
far-infrared (FIR) dust emission maps\footnote{Previous radio/FIR estimates were
$E(B-V) = 0.0080\pm0.0009$ (SBS~1010+535) and $E(B-V) = 0.0460\pm0.0019$ (HS~0747+4259) 
from \citep{Schlegel1998}.  Values from \citep{Schlafly2011} are 14\% smaller.  However, 
based on interstellar gas-to-dust ratios toward AGN at high Galactic 
latitude \citep{Shull2024} we instead used values from the dust optical 
depth map presented in \citep{Planck2016}. This map employed the GNILC technique 
(Generalized Needlet Internal Linear Combination) using spatial information from 
angular power spectra and diffuse component separation to reduce contamination by cosmic 
infrared background radiation.  We multiply the optical depth map at 353 GHz by the 
conversion factor $1.49 \times 10^4$ to obtain $E(B-V)$ in magnitudes.}
from the {\it Planck Observatory}. We adopted $E(B-V) = 0.010$ (SBS~1010+535) 
and $E(B-V) = 0.054$ (HS~0747+4259), both in reasonable agreement with
earlier FIR estimates, although slightly larger.  To apply the reddening to the observed-frame
far-UV, we used the UV selective extinction curve of \citep{Fitzpatrick1999} included in the
Python astropy package.  Shortward of the Lyman edge, in the LLS flux-recovery region, we 
found the true AGN continuum by correcting for the optical depth, 
$\tau_{\lambda} = \sigma_{\lambda} N_{\rm HI}$, of \HI\ photoelectric absorption, 
\begin{equation}
     F_{\lambda}^{\rm (true)} =  F_{\lambda}^{\rm (obs)} \exp [\tau_{\lambda}]   \;  .
\end{equation}
Here, $\sigma_{\nu} \approx \sigma_0 (\nu/\nu_0)^{-3}$ is an approximation to the
\HI\ photoelectric cross section, with $\sigma_0 = 6.304\times10^{-18}~{\rm cm}^{2}$ and
$\nu_0$ defined by the ionization energy $h \nu_0 = 13.598$~eV.
For the actual continuum restoration we use the exact, non-relativistic cross section 
\citep{Bethe1957} with frequency dependence,
\begin{equation} 
   \sigma_{\nu} = \sigma_0 \left( \frac {\nu}{\nu_0} \right)^{-4} 
   \frac {\exp[ 4 - (4 \arctan \epsilon ) / \epsilon ]}  { [1 - \exp(-2 \pi / \epsilon) ] }  \; .
\end{equation}
In this relation, the dimensionless parameter $\epsilon = [(\nu/\nu_0) -1]^{1/2}$.
The two formulae agree at threshold $\nu = \nu_0$, but the approximate formula deviates 
increasingly at shorter wavelengths.  The exact cross section is higher by 8.2\% (700~\AA), 
12.3\% (600~\AA), and 16.4\% (500~\AA), making its inclusion important for EUV continuum restoration.   

\medskip

As flux anchors for fitting the underlying  AGN continua in the EUV, we selected 
windows mostly free of broad emission lines.   These EUV windows were centered 
at $490 \pm 5$~\AA, $530\pm2$~\AA, $575 \pm 5$~\AA, $665 \pm 5$~\AA, $725 \pm 10$~\AA, 
$870 \pm 10$~\AA, 1090--1105~\AA, and 1140--1155~\AA, consistent with earlier AGN composite 
studies \citep{Shull2012,Stevans2014} and shown in light blue in Figures 1 and 2.  
The continuum model is defined as $F_\lambda = F_{\lambda,0} (\lambda/\lambda_0)^{-\alpha_{\lambda}}$, 
where $F_{\lambda,0}$ is the normalization at the (rest-frame) pivot wavelength 
$\lambda_0 = 725$~\AA. To account for the break in spectral slope around 1000~\AA, we 
fitted two indices to the flux distribution in wavelength, designated as $\alpha_{\rm FUV}$ 
in the rest-frame FUV ($\lambda > 1000$~\AA) and $\alpha_{\rm EUV}$ in the 
EUV ($\lambda \leq 1000$~\AA),
\begin{eqnarray}   
F_{\lambda} & \propto & \lambda^{-\alpha_{\rm EUV}} \; \; {\rm for} \; \; \lambda < 1000~\text{\AA} \; \\
F_{\lambda} & \propto & \lambda^{-\alpha_{\rm FUV}} \; \; {\rm for} \; \; \lambda \geq 1000~\text{\AA} \; . 
\end{eqnarray}
These spectral fits were carried out separately, and their extrapolation across the 
transition would lead to a small discontinuity around 1000~\AA.  However, 
this break is non-physical and smaller than the local uncertainty in the fluxes.
For that reason, Figure 1 does not show the continuum across the transition.
The spectral index in the frequency domain was derived from the
fitted $\alpha_{\lambda}$ by the relation $\alpha_{\nu} = (2 - \alpha_{\lambda})$.

\medskip

In the MCMC sampling, we used 50 walkers and 15,000 steps per walker, discarding the first 
10\% steps as burn-in. Convergence was verified by checking the autocorrelation times 
and visual inspection of the posterior distributions. The best-fit broken power-law model 
for the SBS~1010+535 continuum yields a normalization (in relative flux) of 
$F_{\lambda,0} = 1.89\pm0.10$ with spectral indices 
$\alpha_{\rm FUV} = 1.25\pm0.04$ and $\alpha_{\rm EUV} = 0.89\pm0.22$. 
For HS~0747+4259, the best-fit parameters are $F_{\lambda,0} = 1.96\pm0.05$, 
$\alpha_{\rm FUV} = 1.40\pm0.02$, and $\alpha_{\rm EUV} = 1.02\pm0.22$.  
The resulting spectral indices $\alpha_{\lambda}$ and \ion{H}{1} absorber properties 
are summarized in {\bf Table~3}. 

\subsection{Implications for \HeII\ Reionization}  

An AGN spectral index ($\alpha_{\nu} \approx$ 1.7--1.8) between 1--4 Ryd has often been 
adopted to model the large ratios of \HeII\ to \HI\ column densities in the \Lya\ forest
seen toward quasars at $z =$ 2.4--2.9.  The ratios are elevated because \HeII\ is harder 
to ionize (photon energies $E \geq 54.4$~eV) than \HI\ ($E \geq 13.6$~eV), and \HeIII\ 
recombines five times faster than \HII\ at $T \approx 20,000$~K.  The high-redshift AGN 
source spectra are reprocessed and filtered by the intervening IGM, giving some consistency 
with the \HeII\ and \HI\ column densities observed by FUSE and HST.  Their observed ratios, 
denoted by $\eta = N_{\rm HeII}/N_{\rm HI} \approx$ 50--200
\citep{Kriss2001,Fardal1998,Shull2004,Zheng2004,Shull2010}, are considerably greater than the 
primordial abundance ratio by number,  $n_{\rm He}/n_{\rm H} \approx  0.0823$.  

\medskip

The \HeII/\HI\ ratio constrains the relative ionizing fluxes at 912~\AA\ 
(\HI\ continuum) and 228~\AA\ (\HeII\ continuum). In this convention, the specific intensities 
in the ionizing background continua of \HeII\ and \HI\ are expressed as power laws, 
$J_{\nu} = J_i (\nu/\nu_i)^{-\alpha_i}$, with $\alpha_1$ and $\alpha_4$ defined as the spectral 
indices at energies above the ionization thresholds of \HI\ (1~Ryd) and \HeII\ (4~Ryd), 
respectively.  Both $\alpha_1$ and $\alpha_2$ are generally positive, and they
need not be identical. In photoionization ionization equilibrium, when \HII\ and \HeIII\ 
are the dominant ionization stages, we can approximate
$n_{\rm HeIII}/n_{\rm HII} \approx n_{\rm He}/n_{\rm H}$.  Because the \HI\ and \HeII\ 
ionization fractions are usually quite small 
($x_{\rm HI} \ll 1$ and $x_{\rm HeI} \ll x_{\rm HeII} \ll 1$), the ratio of \HeII\ and \HI\ 
can be expressed \citep{Fardal1998,Shull2004,Shull2010} by the approximate formula,
\begin{equation}
    \eta \equiv \frac {N_{\rm HeII}} {N_{\rm HI}} \approx \frac {n_{\rm HeIII}} {n_{\rm HII}} 
           \frac { \alpha_{\rm HeII}^{(A)} } { \alpha_{\rm HI}^{(A)} } 
           \frac {\Gamma_{\rm HI}} {\Gamma_{\rm HeII}} 
         = (1.78) \frac {J_{\rm HI}} {J_{\rm HeII}}
           \frac { (3 + \alpha_4) } { (3 + \alpha_1) } \; T_{4.3}^{0.042} \; .
\end{equation}
For hydrogenic species, with approximate photoionization cross sections 
$\sigma_{\nu} \approx \sigma_i (\nu/\nu_i)^{-3}$ and threshold cross sections 
$\sigma_{\rm HI} = 4 \sigma_{\rm HeII}$, we find photoionization rates  
$\Gamma_i \approx [4 \pi \sigma_i J_i / h (\alpha_i +3)]$.  The potentially 
different indices ($\alpha_1$ and $\alpha_4$) provide minor corrections to the photoionization 
rates of \HI\ and \HeII, and the numerical coefficient (1.78) reflects updated fits to the 
radiative recombination rate coefficients at $T = (10^{4.3}~{\rm K}) T_{4.3}$.  
For the low-density IGM absorbers, we adopt case-A recombination rate coefficients,
$\alpha_{\rm HeII}^{(A)} \approx (1.36\times10^{-12}~{\rm cm}^3~{\rm s}^{-1}) T_{4.3}^{-0.694}$ and
$\alpha_{\rm HeII}^{(A)}  \approx (2.51\times10^{-13}~{\rm cm}^3~{\rm s}^{-1}) T_{4.3}^{-0.736}$,
valid between $T =$10,000--20,000~K \citep{Osterbrock2006}. 

\medskip

In a simplified model of an unprocessed metagalactic background (1--5~Ryd) characterized by 
a flux distribution with a single (typically positive) spectral index 
($\alpha_1 = \alpha_4 \equiv \alpha_b$),
the flux ratio is  $J_{\rm HI} / J_{\rm HeII} = 4^{\alpha_b}$, and the \HeII/\HI\ ratio takes the 
convenient form, 
\begin{equation}
  \eta \approx (1.78) 4^{\alpha_b} \;  .
\end{equation} 
The previously adopted range of AGN spectral indices ($\alpha_{\nu} \approx$ 1.7--1.8) 
would produce fairly low ratios $\eta \approx 20$. A harder ionizing spectrum with 
$\alpha_b \approx$ 1.0--1.1, as found for the  two quasars in this paper, would result 
in an even lower \HeII/\HI\ ratio, $\eta \approx$ 5--6.  However, in both cases, 
reprocessing of the ionizing photons through the IGM softens the background spectrum,
producing the observed range of $\eta =$ 50--200.  The background could also be 
softened by \HeII\ absorption internal to the AGN, a process analyzed in more detail
by \citep{Shull2020}. Unfortunately, there is no easy way to directly observe the AGN 
spectrum at $\lambda \leq 228$~\AA.  The AGN observed by HST/COS only probe down to 
rest-frame wavelengths $\sim350$~\AA.  In addition, at $\lambda \leq 304$~\AA, the 
intergalactic \HeII\ \Lya\ forest will absorb most of the continuum.

\section{Summary and Future Work} 

We now summarize results of our COS observations of two ultra-luminous quasars, 
exhibiting prominent LLS edges at observed wavelengths of 2069~\AA\ (SBS~1010+435) and 
1897~\AA\ (HS~0747+4259).  The moderate optical depths of these LLS, $\tau_{\rm LL} = 1.70\pm0.08$ 
and $1.04\pm0.06$, respectively, allow for significant flux recovery below their Lyman edges 
and provide access to their rest-frame ionizing continuum. The following points summarize our 
primary results:
\begin{enumerate}

\item The rest-frame, de-reddened EUV spectra of two UV-bright, intermediate-redshift quasars  
   were fitted to flux distributions $F_{\nu} \propto \nu^{-\alpha_{\nu}}$, with spectral indices
   $\alpha_{\nu} = 1.11\pm0.22$ (SBS~1010+535 at $z_{\rm AGN} = 1.5086$) and
   $\alpha_{\nu} = 0.98\pm0.22$ (HS~0747+4259 at $z_{\rm AGN} = 1.9006$).
        
\item  As in our previous AGN composite spectra, we see no sign of intrinsic AGN absorption at 
   the ionization edges of \HI\ (912~\AA) or \HeI\ (504~\AA), with optical depth 
   limits $\tau_{\rm HI} < 0.1$ and $\tau_{\rm HeI} < 0.1$.  Evidently, ionizing photons in the 
   AGN ionizing continua of \HI\ and \HeI\ have escape fractions near 100\%. This may not
   be the case for internal \HeII\ absorption, to explain the elevated \HeII/\HI\ ratios
   in the \Lya\ forest at $z =$ 2.4--2.9.

\item Both quasars exhibit prominent \HI\ Lyman Limit systems at redshifts $z_{\rm LLS} = 1.0784$ 
   (SBS~1010+535) and $z_{\rm LLS} = 1.269$ (HS~0747+4259).  With \HI\ column densities 
   $\log N_{\rm HI} =$ 17.43 and 17.22, these LLS have significant flux recovery shortward of
   their Lyman edges, allowing us to observe the underlying EUV continuum (450--912~\AA).
   In the LLS flux-recovery zone of SBS~1010+535, we detect absorption lines in the \HI\ 
   Lyman series and several {$1s\,np \, (^1P) \rightarrow 1s^2 \, (^1S)$} resonance lines of 
   \HeI\ (584, 522, 516~\AA). We also see a possible weak \HeI\ edge (504.26~\AA), appearing 
   at $\lambda \approx 456$~\AA\ in the QSO rest-frame, corresponding to the LLS of \HI. 

\end{enumerate} 

The EUV spectral indices ($\alpha_{\nu} \approx$ 1.0--1.1) of the two AGN in our current paper 
are harder than the mean value $\alpha_{\nu} = 1.41\pm0.15$ in the COS composite spectrum 
\citep{Stevans2014} of 159 UV-bright AGN.  However, most of those AGN had redshifts 
$z_{\rm AGN} < 1.5$, and the sample statistics were poor at rest-frame wavelengths below 600~\AA.  
The spectral indices of the two quasars presented here are similar to the value, 
$\alpha_{\nu} = 0.89 \pm 0.22$ \citep{Tilton2016} in the EUV composite spectrum of 19 
intermediate-redshift AGN.  The redshifts of those quasars ranged from 
$z_{\rm AGN} =$ 0.991--2.142, with 12 of them at $z \geq 1.45$.  
The spectral index and error were derived from a set of Monte-Carlo 
bootstrap-with-replacement experiments among the 19 spectra, with EUV fitting restricted to 
rest-frame wavelengths between 450~\AA\ and 770~\AA.  That spectrum was shown in their Figure 7 
(350--850~\AA) in which the anomalous spectrum of SBS~1010+535 was omitted.  

\medskip

The hard spectra of the two AGN studied here could be related to their extremely high UV luminosities, 
with  $\lambda L_{\lambda} \approx 10^{48}~{\rm erg~s}^{-1}$ at 1350~\AA\ (rest frame).
Both AGN have estimated black-hole masses approaching $10^{10}~M_{\odot}$, with inferred
accretion rates near the Eddington limit.
A harder spectrum could arise from radiative effects on the inner accretion disk at the galactic nucleus.
Alternatively, as suggested in magnetohydrodynamic simulations of accretion disks around
a $10^8~M_{\odot}$ black hole \citep{Jiang2025}, a hard power-law component can appear for photon
energies between 10 eV and 1 keV with a spectral slope varying between $\alpha_{\nu} = 1$ and 2.  
The hard photons are produced above the 
accretion disk by Comptonization within the accretion flow.  This model suggests that a
fraction of ultra-luminous quasars could produce harder EUV spectra for some period of their
accretion history. This effect could produce an observational bias in the  
HST samples used to probe the rest-frame EUV.  In future work, we will explore the spectral
indices in the FUV/NUV for the other 10 AGN in our HST Cycle-30 sample. 

\medskip

To the benefit of future IGM/CGM studies, we are also in a position to measure metal 
ions and perform photoionization models, using the accurate \HI\ column densities in the 
two LLS. The rest-frame EUV can be observed in absorption lines from elements and ions that trace 
high-temperature gas in the halos of two LLS host galaxies.  These include the strong Li-like 
doublets of \NeVIII\ (770.4, 780.3~\AA) and \MgX\ (609.8, 624.9~\AA) and the multi-phase lines of
\OIII\ (833, 702~\AA), \OIV\ (788, 608.4, 554~\AA), \OV\ (630~\AA), \NeIV\ (544~\AA), 
\NeV\ (568~\AA), \NeVI\ (559~\AA), \NIII\ (685~\AA), and \NIV\ (765~\AA).  As shown in 
{\bf Figure 3}, the detection of the strongest EUV absorption lines in the G130M/G160M 
medium-resolution gratings of HST/COS would enable estimates of metallicity and temperature 
in hot halo gas.  

\medskip

In the extended halos of star-forming galaxies, multi-phase gas is frequently seen in \OVI\ 
absorption \citep{Tumlinson2011,Tumlinson2017,Chen2024} out to radial distances of 100--150~kpc. 
Some of this gas may be near the virial temperature of the galaxy, although radiative cooling 
\citep[see e.g.,][]{Bordoloi2017} and dynamical events can trigger gas infall and influence 
galaxy evolution.  The origins and characteristics of LLS have also attracted attention from 
cosmological simulations as examples of galaxy assembly and gas inflow/outflow \citep{Hafen2017}.  
These LLSs connect the IGM, CGM, and galactic halos over the mass range 
$10^{10}$ to $10^{12}~M_{\odot}$. Observations and simulations both suggest that LLS at 
$z \leq 1$ can be associated with galactic winds, tidal interactions in groups, and filamentary 
inflows, all of which produce hot gas at $T \approx 10^6$~K.  
Detections (or non-detections) of the hot phase of halo gas, combined with metallicity estimates, 
would constrain its origin, density, and cooling rate.

\begin{acknowledgements}

 This project was supported by STScI grants in HST~Cycle~30 (HST-GO-17071.001-A) 
 on  ``The Ionizing EUV Continua of Quasars: Minding the Gap" and in an earlier Cycle~25 
 program (HST-GO-15084.001-A) on ``Hot Photons: Measuring the Ionizing Continuum and EUV 
 Emission Lines of Quasars".  We thank the referee for a prompt and careful report,
 which clarified several important issues in our methods and results.  
 We also thank Dr.\ Mark Giroux for helpful discussions of the IGM 
 background and \HeII\ reionization and Dr.\ Yan-Fei Jiang for suggestions about 
 MHD accretion disk models and their emergent spectra.  
  
\end{acknowledgements} 
\vspace{5mm}
\facilities{SDSS}, {HST} 

\software{astropy \citep{Astropy}, rbcodes  \citep{rbcodes}, 
HASP \citep{HASP2024}, emcee \citep{Foreman_Mackey_2013}}

\clearpage

\bibliography{references}{}
\bibliographystyle{aasjournal}
\newpage


\begin{deluxetable} {llccl}
\tablecolumns{5}
 \tabletypesize{\scriptsize}

\tablenum{1}
\tablewidth{0pt}
\tablecaption{Comparison of AGN Composite Spectral Indices\tablenotemark{a} }  

\tablehead{
   \colhead{Paper}
 & \colhead{Survey }
 & \colhead{$N_{\rm AGN}$ } 
 & \colhead{$N_{600}$ } 
 & \colhead{$\alpha_{\nu}$ } 
   }

\startdata
 Zheng \etal\ (1997)   & HST/FOS   & 101 &   4 & $1.96\pm0.15$   \\
 Telfer \etal\ (2002)    & HST/FOS   & 39  & 20  & $1.57\pm0.17$  \\
 Scott \etal\ (2004)     & FUSE      & 85  &  0  & $0.56^{+0.38}_{-0.28}$  \\
 Shull \etal\ (2012)     & HST/COS   & 22  &  6  & $1.41 \pm 0.21$    \\
 Stevans \etal\ (2014) & HST/COS   & 159 & 10  & $1.41 \pm 0.15$   \\
 Lusso \etal\ (2015)    & HST/WFC3  &  53 &  0  & $1.70\pm0.61$     \\
 Tilton \etal\ (2016)     & HST/COS   & 19  & 19  & $0.89\pm0.22$    \\
                                   &           &     &     &                  \\
{\bf Current Results:} &           &     &     &                   \\
SBS~1010+5345       & HST/COS   &  1  &  1  & $1.11\pm0.22$    \\
HS~0747+4259         & HST/COS   &  1  &  1  & $0.98\pm0.22$    \\                     
\enddata 

\tablenotetext{a} {Previous EUV composite spectra of AGN with spectral indices, 
$\alpha_{\nu}$ were based on varying numbers of targets $N_{\rm AGN}$.  Only 
a few ($N_{600}$) extended to rest-frame wavelengths shortward of 600~\AA.
The final two lines show results for the two AGN in the current paper. } 

\end{deluxetable}



\begin{deluxetable} {llll clll}
\tablecolumns{8}
 \tabletypesize{\scriptsize}

\tablenum{2}
\tablewidth{0pt}
\tablecaption{HST/COS Observations\tablenotemark{a} }  

\tablehead{
   \colhead{Target}
 & \colhead{R.A.}
 & \colhead{Decl.} 
 & \colhead{Grating} 
 & \colhead {Center}
 & \colhead{Exp.\ Time}
 & \colhead{Exp.\ Date} 
 & \colhead{Program ID} 
 \\
 \colhead{  }
 & \colhead{(J2000)}
 & \colhead{(J2000)} 
 & \colhead{ } 
 & \colhead{(\AA)} 
 & \colhead{(s)}
 & \colhead {(GMT)} 
 & \colhead{(and P.I.)} 
 }

\startdata
SBS~1010+535 & 153.3756767 & +53.266561 & G230L & 2365 & 2369.088 & 2023 Apr 15 & GO-17071 (Shull) \\
SBS~1010+535 & 153.3756867 & +53.266561 & G230L & 3000 & 2596.768 & 2023 Apr 15 & GO-17071 (Shull)  \\
SBS~1010+535 & 153.3756867 & +53.266561 & G230L & 3360 & 2596.736 & 2023 Apr 15 & GO-17071 (Shull) \\
SBS~1010+535 & 153.3756867 & +53.266561 & G140L & 1105 & 5460.704 & 2013 Oct 1   & GO-13302 (Shull) \\
SBS~1010+535 & 153.3756867 & +53.266561 & G140L & 1105 & 5460.736 & 2014 Jan 30 & GO-13302 (Shull)  \\
                           &                      &                     &             &          &                 &                      &                             \\
HS~0747+4259 & 117.7276992 & +42.872008 & G230L & 2365 & 2273.088 & 2023 Feb 17 & GO-17071 (Shull) \\
HS~0747+4259 & 117.7276992 & +42.872008 & G230L & 3000 & 2500.832 & 2023 Feb 17 & GO-17071 (Shull) \\
HS~0747+4259 & 117.7276992 & +42.872008 & G230L & 3360 & 2500.800 & 2023 Feb 17 & GO-17071 (Shull) \\
HS~0747+4259 & 117.7276992 & +42.872008 & G140L & 1105 & 7973.120 & 2018 May 24 & GO-15084 (Shull) \\
\enddata 

\tablenotetext{a} {Both AGN targets were observed with low-resolution COS gratings in the 
far-UV (G140L) in a single central wavelength setting (1105~\AA).  In the near-UV (G230L) 
we used three wavelength settings (2365, 3000, and 3360~\AA).} 
 
\end{deluxetable}



\begin{deluxetable} {lll llll}
\tablecolumns{7}
 \tabletypesize{\scriptsize}

\tablenum{3}
\tablewidth{0pt}
\tablecaption{Fits to AGN Continuum and LLS\tablenotemark{a} }  

\tablehead{
   \colhead{Target}
 & \colhead{$z_{\rm AGN}$ }
 & \colhead{$z_{\rm LLS}$ } 
 & \colhead{$\log N_{\rm HI}$ }
 & \colhead{E(B-V)}  
 & \colhead{$\alpha_{\rm FUV}$ } 
 & \colhead {$\alpha_{\rm EUV}$}
   }

\startdata
SBS~1010+535 & 1.5086 & 1.269  & $17.431\pm0.021$ & 0.010 & $1.25\pm0.04$ & $0.89\pm0.22$   \\
HS~0747+4259 & 1.9006 & 1.0784 & $17.218\pm0.024$ & 0.054 & $1.40\pm0.02$ & $1.02\pm0.22$        
\enddata 

\tablenotetext{a} {Fits to the underlying AGN continuum with power-law forms,
$F_{\lambda} \propto \lambda^{-\alpha}$, with different indices for 
the rest-frame far-UV ($\alpha_{\rm FUV}$ at $\lambda > 1000$~\AA) and rest-frame
EUV ($\alpha_{\rm EUV}$ at $\lambda < 1000$~\AA).  The de-reddened continua were 
corrected for LLS at redshifts $z_{\rm LLS}$ and column densities 
$N_{\rm HI}$ (cm$^{-2}$). The corresponding spectral indices in the frequency domain
are $\alpha_{\nu} = (2 - \alpha_{\lambda})$. Thus, we find EUV indices of 
$\alpha_{\nu} = 1.11\pm0.22$ (SBS~1010+435) and $0.98\pm0.22$ (HS~0747+4259). }

\end{deluxetable}



\begin{figure}[ht]

\centering
\includegraphics[angle=0,scale=0.25]{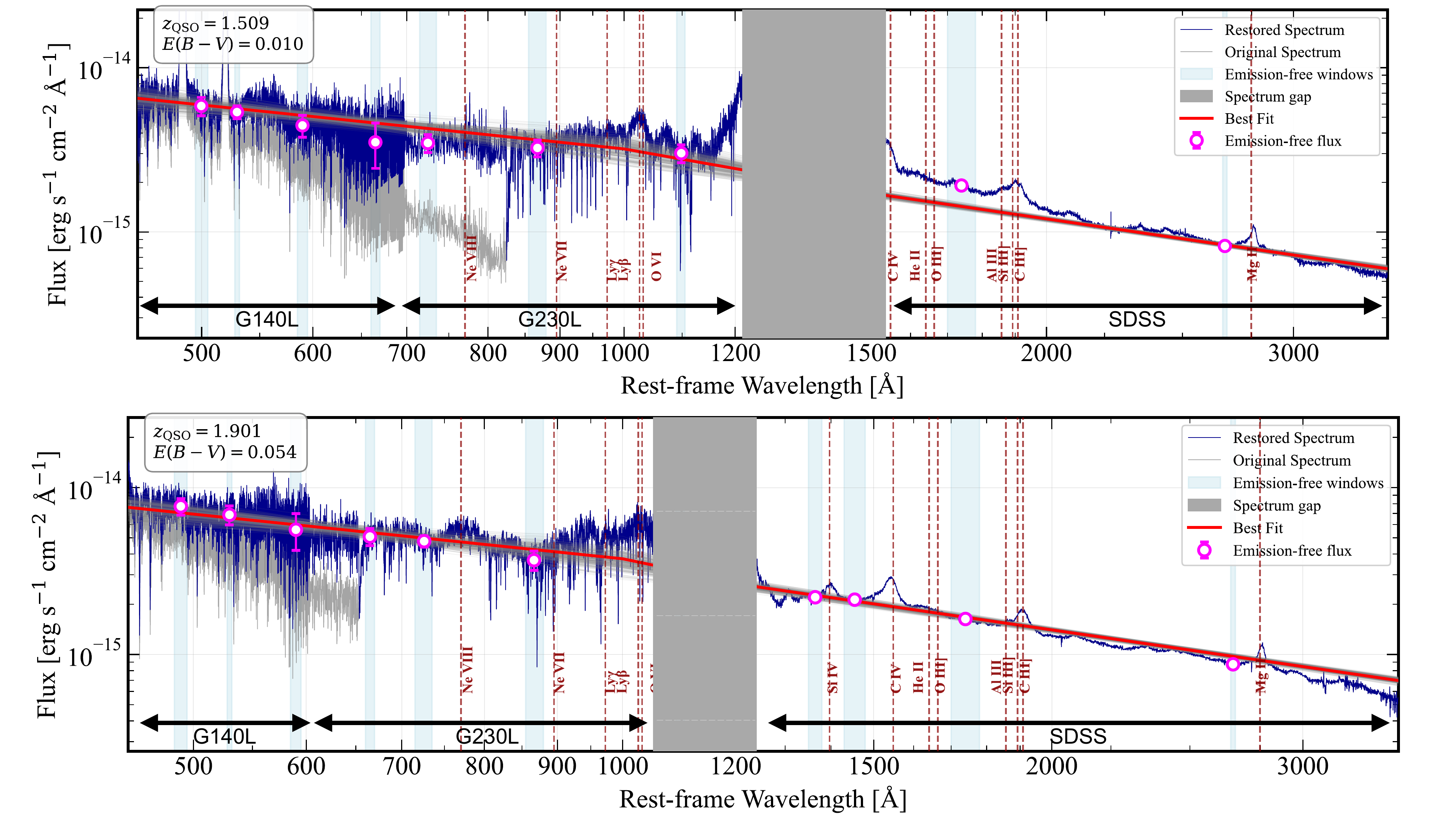}
\caption{Combined far-UV (FUV), near-UV (NUV), and optical spectra of two 
intermediate-redshift quasars, SBS 1010+535 ($z = 1.9006$) and HS 0747+5259 
($z = 1.5086$), obtained with HST/COS (G140L and G230L gratings) and SDSS, 
respectively. The HST/COS spectra are co-added separately for each grating 
using the HASP pipeline and stitched together via inverse-variance weighting. 
The SDSS optical spectra are flux-scaled by factors of 0.80 (SBS~1010) and 0.75 
(HS~0747) to match the COS flux level. The original observed spectra are shown 
in gray, while the de-reddened and continuum-restored spectra are overplotted 
in dark blue. Prominent quasar emission lines are marked by vertical dashed lines, 
and our continuum-fitting windows (emission-free regions) are highlighted with 
light blue bands. The gray vertical band marks the gap between the HST and 
SDSS wavelength coverage. In SBS~1010+535, the geocoronal emission lines 
of \Lya\ (1216~\AA) and \OI\ (1304~\AA) appearing at 485~\AA\ and 520~\AA\ 
(quasar rest-frame) were excised from the fitting. }
\label{Figure:spectra}

\end{figure}
 


\begin{figure}[h]
\centering
 \includegraphics[angle=0,width=16cm] {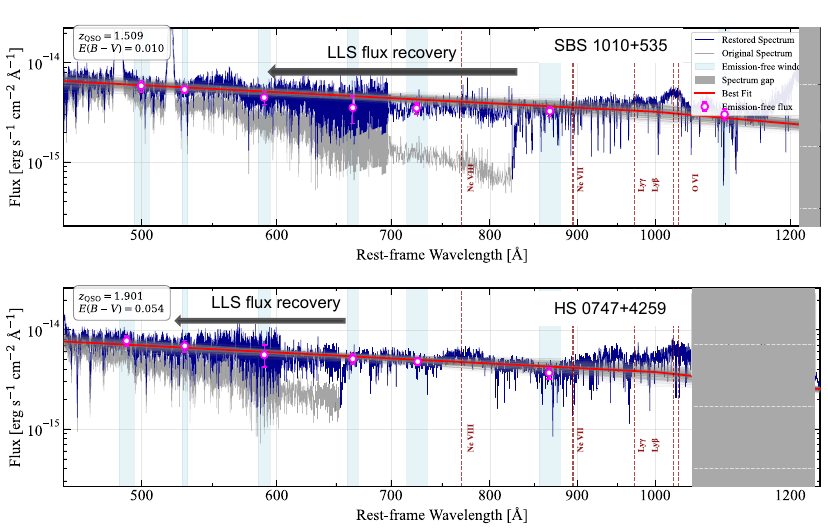}  
 \caption{Low-resolution HST/COS spectra toward two bright AGN, plotted in their 
 rest frame and showing the EUV flux recovery shortward of two intermediate redshift LLSs.  
{\bf (Top)} SBS~1010+535 ($z_{\rm AGN} = 1.5086$) with LLS at $z_a = 1.269$ 
and optical depth $\tau_{\rm LL} = 1.70\pm0.08$. 
{\bf (Bottom)}  HS~0747+4259 ($z_{\rm AGN} = 1.9006$) with LLS at $z_a = 1.0784$
and optical depth $\tau_{\rm LL} = 1.04\pm0.06$.  Absorption-corrected EUV continua 
(in dark blue) were obtained by multiplying the observed flux (in grey) by 
$\exp(\tau_{\lambda})$ (see eqs.\ [3] and [4]) to account for \HI\ absorption.  
The AGN continuum in the EUV was fitted through five wavelength bands (light blue zones) 
generally free of strong AGN emission lines (M.\ Stevans \etal\ 2014).  Two other 
line-free windows (see Section 2.2) contribute to the rest-frame FUV fits. 
The higher noise level at shorter wavelengths arises from the transition in 
gratings from G230L to G140L and the smaller G140L pixel size. }
\label{Figure:contfit}
\end{figure}



\begin{figure}[h] 
\centering
 \includegraphics[angle=0,width=18cm] {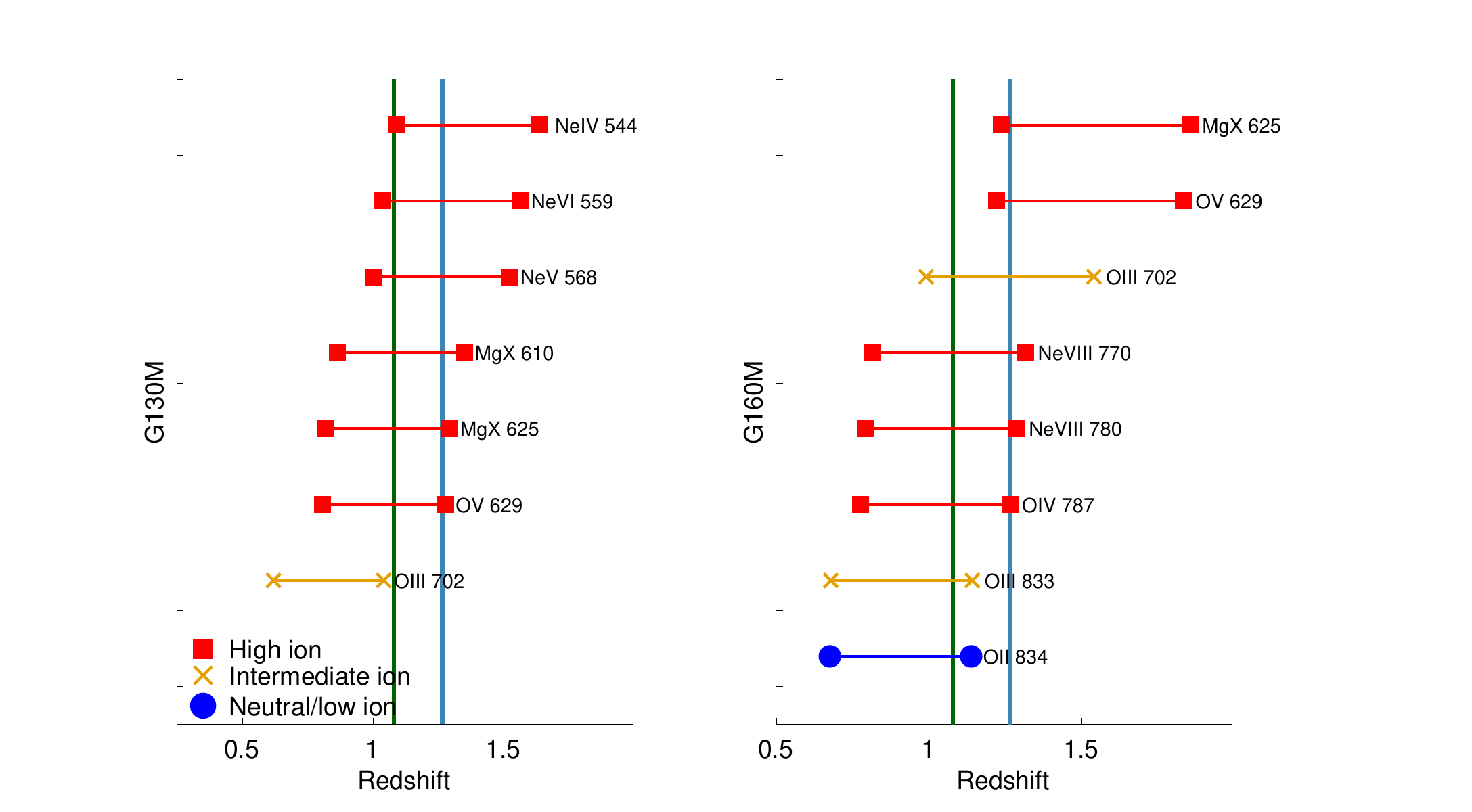}  
 \caption{Locations of various far-UV and EUV absorption lines from metal ions, with  
 ranges of COS observability (G130M and G160M).  Redshifts of the two LLS are shown as 
 vertical lines at $z_{\rm LLS} = 1.0784$ and 1.269. The Li-like doublets (\MgX, \NeVIII) 
 are observable in both LLS, together with high ionization lines of O and Ne. }
 
\end{figure}


\end{document}